# Oscillometric Blood Pressure Measurement Using a Hybrid Deep Morpho-Temporal Representation Learning Framework

Niloufar Delfan and Mohamad Forouzanfar, Senior Member, IEEE

*Abstract*— *Objective*: Oscillometric monitors are the most common automated blood pressure (BP) measurement devices used in non-specialist settings. However, their accuracy and reliability vary under different settings and for different age groups and health conditions. A main limitation of the existing oscillometric monitors is their underlying analysis algorithms that are unable to fully capture the BP information encoded in the pattern of the recorded oscillometric pulses. *Methods*: In this paper, we propose a new 2D oscillometric data representation that enables a full characterization of arterial system and empowers the application of deep learning to extract the most informative features correlated with BP. A hybrid convolutional-recurrent neural network was developed to capture the oscillometric pulses morphological information as well as their temporal evolution over the cuff deflation period from the 2D structure, and estimate BP. *Results*: The performance of the proposed method was verified on three oscillometric databases collected from the wrist and upper arms of 245 individuals. It was found that it achieves a mean error and a standard deviation of error of as low as 0.08 mmHg and 2.4 mmHg in the estimation of systolic BP, and 0.04 mmHg and 2.2 mmHg in the estimation of diastolic BP, respectively. *Conclusion*: Our proposed method outperformed the state-of-the-art techniques and satisfied the current international standards for BP monitors by a wide margin. *Significance*: The proposed method shows promise toward robust and objective BP estimation in a variety of patients and monitoring situations.

## I. INTRODUCTION

BLOOD pressure (BP) is an important vital sign and plays a key role in comprehending the hemodynamic condition of an individual [4, 5]. High BP is a common chronic disease and represents the primary risk factor for stroke and heart disease, which are leading causes of death in the United States [6]. High BP affects 1.13 billion people worldwide [8], but only about 1 in 4 adults with hypertension monitor their condition and keep it under control.

Among non-invasive blood pressure measurement techniques, oscillometric method offers the best opportunity for automation and provides more accurate results. It can be repeatedly performed by users at home and operated in a noisy environment [10]. Oscillometry can be seen as a system with three main components including the artery, the cuff, and the arm. The inputs of the system are the applied cuff pressure and the intra-arterial blood pressure. As the cuff is inflated to a supra-systolic pressure, the arterial lumen area decreases until it becomes occluded. The cuff is then deflated gradually to a sub-diastolic pressure while recording the pressure variations within the cuff. The measured waveform, commonly known as cuff deflation curve, is composed of two main components: the slow-varying component representing the applied cuff pressure and the pulsations that are caused by the intra-arterial pressure through the cuff-arm-artery interaction. These pulsations are extracted using a high-pass filter to form a signal known as the oscillometric waveform (OMW) [10]. Fig. 1(a) illustrates the oscillometry physical setup.

Many automatic algorithms have been developed to estimate systolic BP (SBP), diastolic BP (DBP), and mean arterial pressure (MAP) from OMW [7, 10, 12-15]. Most of the existing oscillometric algorithms rely on processing the OMW envelope ignoring the wealth of information contained in the morphology of individual oscillometric pulses at different cuff pressures [2, 3, 14, 16, 17]. Recently, several feature engineering algorithms have been proposed to characterize the morphological changes of individual oscillometric pulses and to estimate BP thereof [12, 13]. However, these extracted features are user defined and may not fully capture all the BP-correlated morphological information contained in the individual oscillometric pulses. Moreover, the evolution of these features as a function of cuff pressure contains valuable information about the arterial system behavior, including BP, that has mostly been ignored. A fully automatic characterization of oscillometric data is therefore required that can completely capture the morphological information of individual oscillometric pulses and their evolution as a function of cuff pressure. This comprehensive characterization is a prerequisite to the development of advance intelligent techniques for estimation of BP.

In this paper, we propose a novel oscillometric data representation approach that can capture the entire information contained in the oscillometric pulses morphology and model their varying temporal pattern and relationships over cuff deflation. The approach is based on assigning each oscillometric pulse to its corresponding cuff pressure in a two-dimensional (2D) array structure where the x-axis represents the cuff pressure (or time), and the y-axis represents the

| TABLE I |||||
|---|---|---|---|
| CHARACTERISTICS OF INDIVIDUALS IN DIFFERENT DATASETS |||||
| Characteristics | Dataset1 | Dataset2 | Dataset3 |
|---|---|---|---|
| Number of individuals | 85 | 150 | 10 |
| Records per individual | 5 | 1-5 | 15 |
| SBP: Min, Max [mmHg] | 78, 147 | 80, 199 | 79, 136 |
| DBP: Min, Max [mmHg] | 42, 99 | 36, 104 | 52, 86 |
| Age: Min, Max | 12, 80 | 23, 97 | 24, 63 |
| Sex: Male, Female | 48, 37 | 82, 68 | 6, 4 |

sample index of individual pulses (or morphology). To handle missing pulses of the varied-length recorded OMWs, inter- and extrapolation techniques were employed to reconstruct an OMW over a predefined cuff pressure interval. To avoid hand-crafted engineered features, we employed deep learning, the new family of machine learning algorithms that can fully automate learning process from well-structed raw data. A deep hybrid convolutional (CNN) recurrent neural network (RNN) was developed to extract the most relevant morphological and temporal information contained in our 2D array and estimate BP. Our new approach was thoroughly evaluated on three different oscillometric datasets obtained from upper arm and wrist and was compared with the state-of-the-art oscillometric BP measurement techniques.

## II. METHODOLOGY

### A. Study Participants

This study was conducted on three separate datasets collected under different settings. Table I summarizes the participants characteristics in each dataset.

The first dataset was provided by Biosign Technologies Inc., Toronto, Canada. It was acquired using an automated wrist BP monitor in accordance with the recommendations of the American National Standards Institute (ANSI)\Association for the Advancement of Medical Instrumentation (AAMI) SP10 standard [19]. The dataset includes five sets of oscillometric wrist BP measurements obtained from 85 individuals. The average values of two independent nurse measurements obtained with a double stethoscope (following each oscillometric recording) is used as the reference. This study was confirmed by a local research ethics committee, and written informed consent was obtained from all participants.

The second dataset has been published by the University of New South Wales, Sydney, Australia [20]. It includes 350 upper arm records obtained from 150 individuals collected using a multi-parameter clinical monitoring unit. The reference BP values were obtained by automatic signal processing of simultaneously measured Korotkoff sounds as described in [21]. Ethics Approval from the University of New South Wales was obtained to perform this study.

The third dataset was collected at the University of Ottawa, Ottawa, Canada. It consists of 150 upper arm recordings obtained from 10 healthy individuals. Food and Drug Administration (FDA)-approved Omron monitor (HEM-790IT) BP readings obtained from the opposite arm were used as the reference [23]. This study was approved by the University of Ottawa Research Ethics Board, and written informed consent was obtained from all participants.

### B. Problem Formulation

We formulated the problem of BP estimation from a varied-length OMW into a deep neural network framework by representing the oscillometric pulses based on their corresponding CP. The model input is a 2D array with a fixed size generated by inter- and extrapolation of original oscillometric pulses, and its output is a single value, indicating SBP or DBP. The model's objective is to minimize the mean square error (MSE) between the reference labels and outputs.

### C. 2D Morpho-Temporal Oscillometric Data Representation

To fully characterize the variations in oscillometric pulse morphology over the cuff deflation period, a 2D structure was created (see Fig. 1) where each column represents the oscillometric pulse morphology at a specific cuff pressure.

To create this 2D structure, a 4th order low-pass digital Butterworth filter with a cutoff frequency of 10 Hz was first applied to reduce high-frequency noise and artifacts.

Next, oscillometric peaks were detected using the automatic multiscale-based peak detection (AMPD) algorithm with an adaptive sliding window [24]. To find the troughs, the minimum value in the group of points between two consecutive peaks was first detected. By denoting the amplitude of the i-th peak as $P_i$ and the amplitude of the i-th minimum as $M_i$, a threshold for each pair of minimum and peak was defined as thr = $4\times(P_{i+1}-M_i)/5$. A trough $T_i$ was detected between the i-th minimum and i-th peak as the closest sample to the peak with the lowest amplitude among its previous and next five samples (local minimum) while not reaching $P_{i+1}$-thr. If no samples met the above criteria, the minimum $M_i$ was chosen as the trough. The oscillometric pulses were then extracted as those samples between two consecutive troughs.

Outliers were detected and removed by the analysis of pulse duration and amplitude. The median pulse duration ($MED_d$) was calculated for each OMW and those pulses whose duration was out of $MED_d \pm 0.3$ sec range were detected as outliers. To detect outliers based on the pulse amplitude, a modified z-sore was defined as MZ-score = $0.6745\times (A_i - \bar{A})/MAD_A$, where $A_i$ is the amplitude of the i-th pulse ($P_i$-$T_i$), $\bar{A}$ is the average pulse amplitude, and $MAD_A$ is the median absolute deviation of pulse amplitudes. Those pulses with a MZ-score of greater than 10 were identified as outliers (see Fig. 1(b)). As differences in the scales of input variables may impact the performance of any machine learning algorithm, each OMW was normalized using a maximal absolute scaling such that the maximal absolute amplitude of pulses in the OMW was mapped to 1.

The OMWs were then reshaped into a 2D structure where x axis represented the cuff pressure evenly sampled at 1 mmHg from the minimum (20 mmHg) to maximum (235 mmHg) existing values in our datasets (leading to 215 indices). Each extracted oscillometric pulse was resampled at 215 samples/pulse and positioned in the column corresponding to the closest pressure at which the oscillometric pulse occurred (see Fig. 1(c)). Given that the duration of each pulse is around 1 sec, 215 samples per pulse will fully capture all the pulse morphology features while making our 2D structure a square matrix of 215 by 215 samples. Given that some columns of the 2D structure may not contain any pulses (due to outlier removal, fast deflation, or the different cuff pressure ranges), all the missing data were replaced by performing linear inter- and extrapolation.

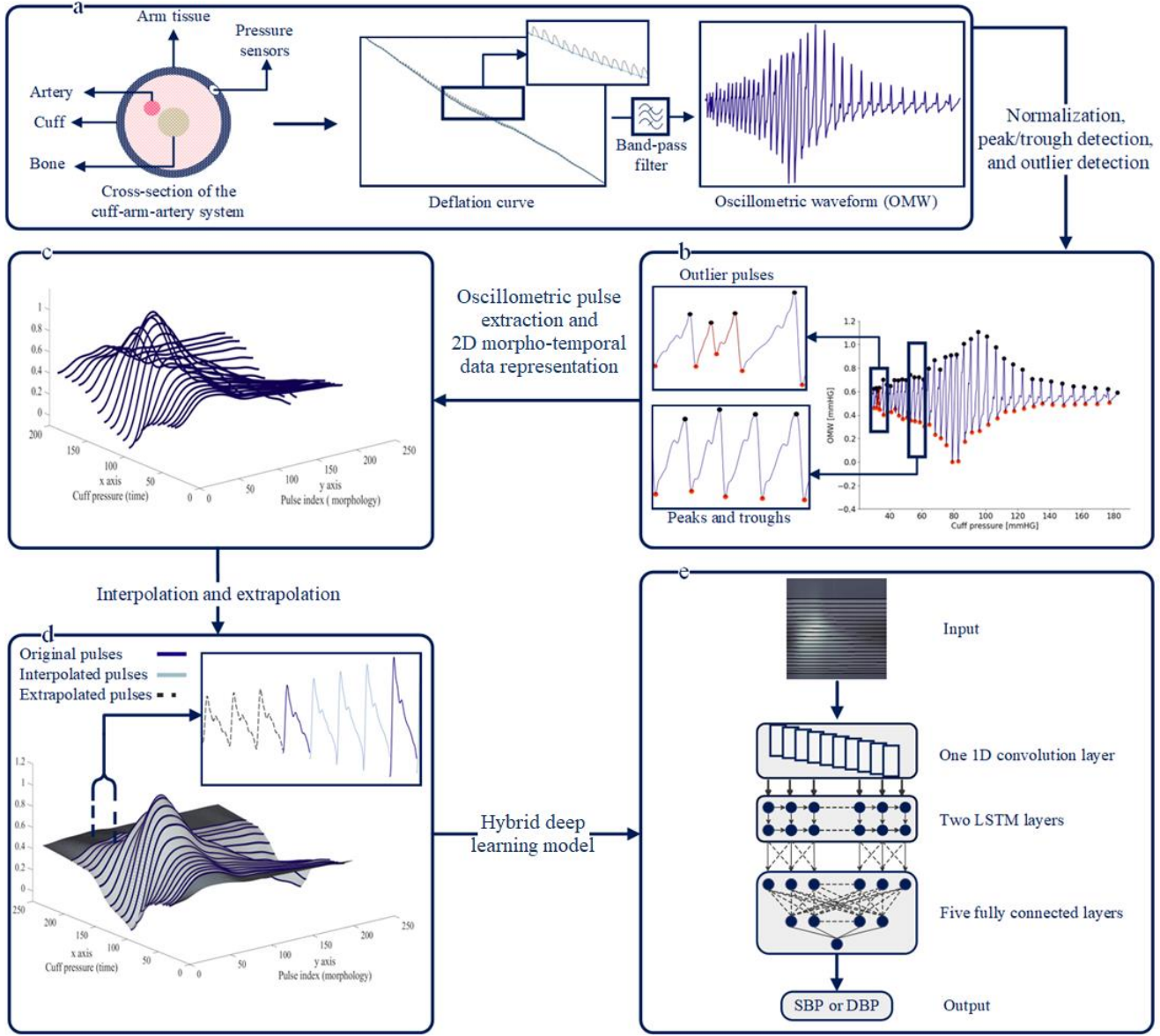

Fig. 1. Architecture of our proposed hybrid deep morpho-temporal representation learning framework for oscillometric BP estimation. (a) Oscillometry system physical setup. (b) the oscillometric waveform extracted from the cuff deflation curve. The detected peaks and troughs are shown by black and red circles, respectively. The detected outlier pulses are shown in red. (c-d) The proposed 2D morpho-temporal oscillometric data representation before and after inter- and extrapolation. Blue lines show the original pulses, light gray lines show the interpolated pulses, and dark gray lines show the extrapolated pulses. (e) Architecture of our proposed hybrid CNN-RNN model.

### D. Hybrid Deep Learning Model

Our proposed deep learning model was constructed by integrating a 1D convolutional neural network (1D-CNN), two long short-term memory (LSTM) layers, and five fully connected layers (see Fig. 1(e)).

A CNN was developed to capture the oscillometric pulses morphological features. The CNN model consisted of 10 kernels of width 107 applied in the direction of cuff pressure (x-axis). The kernels extracted and fused pulse morphology information leading to a 10 (oscillometric pulse feature map) by 215 (representing the evolution of the feature map over cuff deflation) array that was fed to two LSTM layers. The stride was set to 1 and boundary padding was set to the half of the kernel size to produce output of the same size as the input. To prevent vanishing gradient problem, a rectified linear unit (ReLU) was used in the convolutional layer.

Two LSTM layers were used to model the evolution of morphological features extracted from oscillometric pulses over different cuff pressures. Each layer received 215 cuff pressure steps (timesteps) of dimension 10. Through training the parameters of the LSTM cells, the network learned to remember or forget certain parts of the information in the series of oscillometric pulses and extract a morphological-temporal feature map. The number of hidden units was set to 10 leading to an output array of 215 by 10 which was flattened and fed into the fully connected layers.

The fully connected module assigned weights to the extracted morphological-temporal features in five nonlinear hidden layers with 1000, 500, 250, 100, and 50 nodes and a linear output layer with a single node. The hidden layer's activation functions were ReLU that empowered the network to learn the complex and nonlinear relationship between the inputs and the targets. The linear output layer made it possible to have outputs of any range facilitating the estimation of BP.

*E. Network Training and Model Validation*

A subject-independent test (SIT) was performed to evaluate and compare the performance of our proposed method [25]. In SIT, data are separated so that all samples belonging to each individual are included only in the training or the test set. Since in real-world settings it is difficult to obtain information about a new test subject in advance, SIT provides a fair evaluation of the model's performance when it is employed on new individuals. To this end, a leave-one-subject-out cross-validation was performed to train, validate, and test our proposed model where each dataset was divided into two sets, called train-validation and testing sets. In each evaluation fold, all recordings belonging to an individual were selected for testing, and the rest of the data were used for training and validation. Using this procedure, we produced the largest possible training and validation set, while testing the model on unseen test data. One record from each individual in the training set was randomly chosen to create a validation set (used to tune hyperparameters). This procedure was then repeated so that the data belonging to each induvial was once placed in the test set. All the reported results are based on the test set.

The process of determining the optimal model parameters was performed by minimizing a loss function defined as the MSE between the network estimated BP and the target values. The minimization was performed by calculating the gradient using the backpropagation technique in a batch style. The batch size was set to size of the training set. To overcome overfitting, L1 regularization technique with the default regularization factor $Λ = 0.0001$ was employed. Note that the biases were not regularized. Also, the early-stopping technique was utilized for the entire model. The networks were trained with the training data, and the validation error was checked at every iteration. When the validation error increased for a specified number of successive iterations (set to 30), the networks' parameters that lead to the best validation performance were saved. A plateau learning rate schedule was utilized where the learning rate was automatically reduced once the model performance declined for some epochs. The initial learning rate was set to 0.001 and the patience for the reduction of learning rate was set to 10.

III. RESULTS AND DISCUSSION

*A. Performance Evaluation*

Three datasets were used to evaluate the generalization of our deep-learning model (see Table I). Mean error (ME), mean absolute error (MAE), and standard deviation of the error (SDE) were used to evaluate the algorithms' performance. To reduce randomness, these values were averaged over ten runs. The results were also validated against the British Hypertension Society (BHS) [26] and ANSI/AAMI/ISO [19] standards for automated BP monitors. In addition, Bland-Altman plots were used to evaluate the performance of our proposed technique at different BP values.

Table II summarizes the performance of our proposed method on three oscillometric datasets in terms of ME, MAE, and SDE obtained on the test set. ME, shows the overall bias of the algorithm which is negligible in most cases ranging from 0.08 mmHg to 0.22 mmHg in the estimation of SBP and from 0.00 mmHg to 0.11 mmHg in the estimation of DBP. SDE shows the error variability which ranged from 2.48 mmHg to 3.71 mmHg and 2.22 mmHg to 3.45 mmHg in the estimation of SBP and DBP, respectively. MAE shows an overall performance evaluation of the algorithm by considering both the bias and the variance. The MAE ranged from 1.88 mmHg to 2.98 mmHg, and 1.65 mmHg to 2.30 mmHg in estimation of SBP and DBP, respectively. It is observed that all the error metrics are relatively low on all the three datasets. The highest ME, SDE, and MAE were observed for the estimation of SBP in Dataset 3. This can be related to the smaller size of this dataset compared to the other two which led to a lower learning performance. It was also observed that the proposed method meets the ANSI/AAMI/ISO standard for automated BP monitor at its bias (ME) and SDE were substantially lower that 5mmHg and 8mmHg, respectively. It should be noted that only our first dataset fully met the patient's population requirements for the ANSI/AAMI/ISO standard.

The proposed method was also validated against the BHS protocol of automated BP monitors. According to the BHS standard, if 60% of a device's measurements are within 5 mmHg, 85% are within 10 mmHg, and 95% are within 15 mmHg of the gold standard mercury measurements, the device is considered Grade A. Accordingly, Grades B, C, and D are defined with more relax criteria [26]. Table III reports the performance of the proposed method with respect to BHS standard criteria. It is observed that it achieves the highest BHS ranking (grade A) in the estimation of SBP and DBP on all the three datasets.

Bland-Altman plots were employed to compare the estimates obtained by our proposed method and the reference SBP and DBP. Fig. 2 shows the distribution of estimation errors. The x-axis represents the average estimates of our proposed method and the reference measurements, and the y-axis represents the difference between the two methods. It is observed that in most cases the error is uniformly distributed along different BP values. A slight pattern of underestimation of high DBP values and overestimation of low SBP and DBP values is observed for Dataset 2. Given that such samples were rare in our dataset, the method requires more training samples in these ranges of BP to be able to correctly estimate them. In overall, SBP and DBP estimates obtained by our proposed model are in close agreement with the reference readings.

TABLE II
PERFORMANCE OF THE PROPOSED METHOD IN ESTIMATION OF SBP AND DBP REPORTED ON UNSEEN TEST DATA (ERRORS ARE IN MMHG).

|  | Dataset1 | | | Dataset2 | | | Dataset3 | | |
| --- | --- | --- | --- | --- | --- | --- | --- | --- | --- |
|  | SDE | ME | MAE | SDE | ME | MAE | SDE | ME | MAE |
| SBP | 2.48 | 0.08 | 1.88 | 3.65 | 0.16 | 2.07 | 3.71 | 0.21 | 2.98 |
| DBP | 2.22 | 0.04 | 1.65 | 3.45 | 0.10 | 2.30 | 2.25 | 0.00 | 1.78 |

TABLE III
BHS GRADING OF THE PROPOSED METHOD.

|  | Dataset1 | | | | Dataset2 | | | | Dataset3 | | | |
| --- | --- | --- | --- | --- | --- | --- | --- | --- | --- | --- | --- | --- |
|  | BHS Grade | <=5 | <=10 | <=15 | BHS Grade | <=5 | <=10 | <=15 | BHS Grade | <=5 | <=10 | <=15 |
| SBP | A | 89.62 | 99.13 | 99.71 | A | 90.28 | 98.00 | 98.85 | A | 84.00 | 98.00 | 100 |
| DBP | A | 95.29 | 100 | 100 | A | 86.00 | 98.28 | 99.71 | A | 98.00 | 100 | 100 |

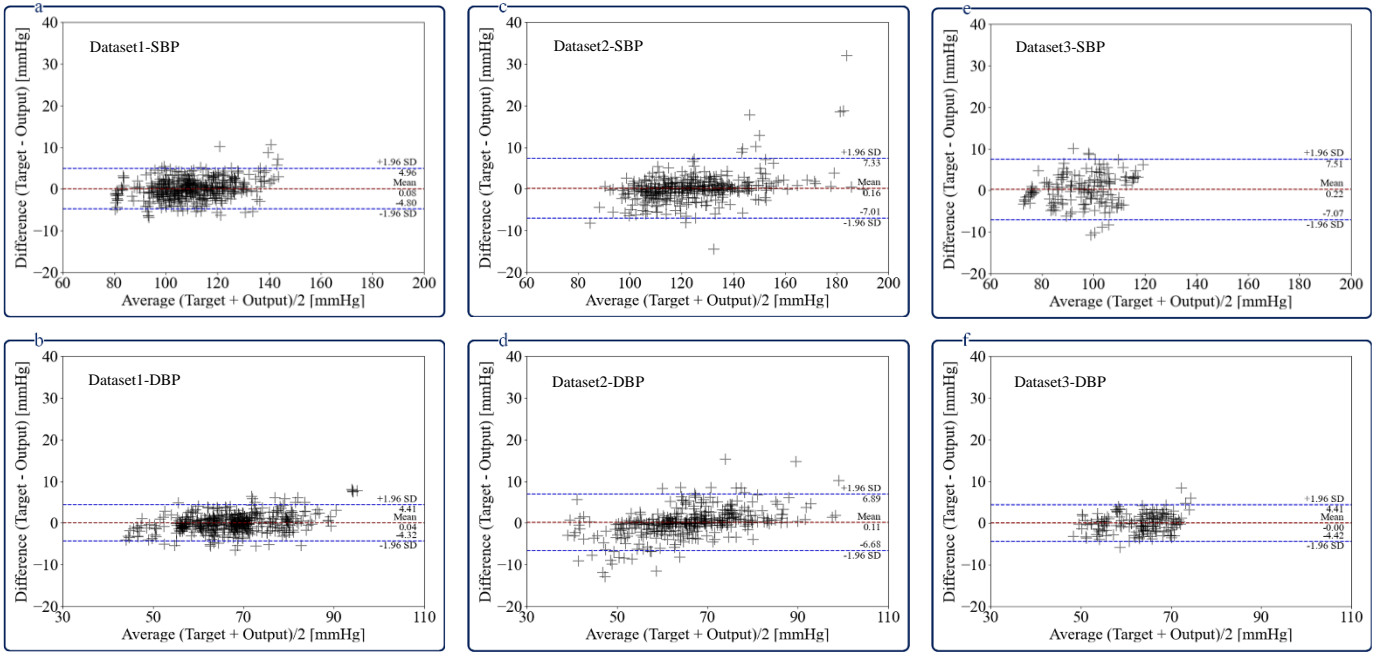

Fig. 2. Bland-Altman plots showing the performance of the proposed method in the estimation of SBP and DBP on three different datasets.

*B. Hyperparameters Selection*

Our proposed model parameters were learned during training from the available data, and its hyperparameters were tuned to optimize its fit. Several experiments were performed to tune the hyperparameters and examine the performance of our algorithm. Firstly, the number of training epochs was tuned. It was observed that the performance of each fold improved dramatically until about 100 epochs and fluctuated constantly when the epoch number was increased. Next, different number of hidden LSTM units and CNN feature channels were examined (5 to 215). It was observed that (see Fig. 3) the best performance is achieved with 10 hidden units and 10 CNN feature channels (kernels). An over simplified model underfitted the data while a more complex network led to overfitting. Same procedure was also followed to find the optimum learning rate, number of CNN layers, number of LSTM layers, and the dimension of convolution kernels as reported in Section II.E.

*C. Ablation Study*

To further evaluate the contribution of different modules of our proposed method on its overall performance, an ablation study was performed where the proposed method was evaluated using only a CNN model, and CNN+LSTM model, and a CNN+2LSTM model (our final model). The results achieved on different datasets are reported in Table IV. It was observed that the SBP and DBP SDEs were improved on average (averaged over all datasets) by 3.1 mmHg, and 3.2 mmHg when adding an LSTM layer to the CNN module, and by 1.5 mmHg and 1.5 mmHg when adding the second LSTM layer. Similar improvements were observed in terms of MAE. No further improvement was achieved adding more LSTM layers.

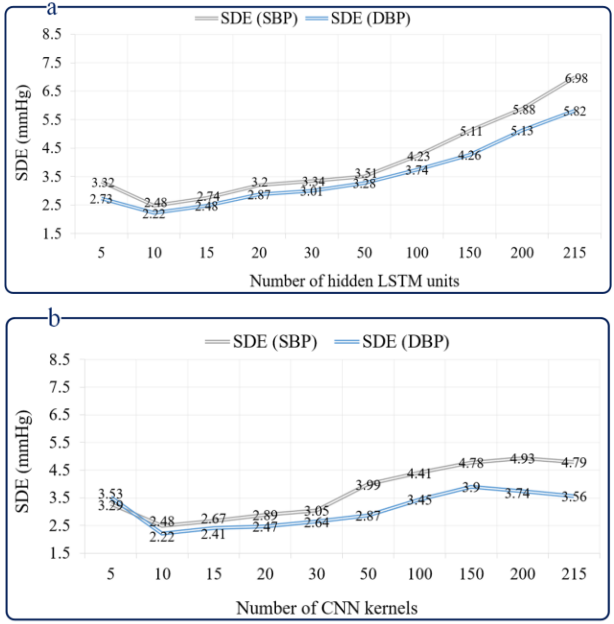

Fig. 3. Summary of the SBP and DBP SDE obtained on Dataset 1 using (a) different numbers of LSTM units, (b) different numbers of CNN kernels.

*D. Comparison with the State-of-the-Art*

A comparison of the proposed approach with other conventional and state-of-the-art oscillometric BP estimation techniques is provided in Table V. These techniques include maximum amplitude algorithm (MAA) [1, 12], feed-forward NN (FFNN) [2, 12], support vector regression (SVR) [3, 12], Bayesian fusion [22], deep belief networks (DBN)-deep neural networks (DNN) [3] and its extended ensemble version [7] deep Boltzmann regression (DBR) [9], Dempster–Shafer fusion based on a deep Boltzmann machine (DBMDS) [11], LSTM-RNN [12], pulse transit time (PTT) [13], and ratio-independent modeling [16]. It is observed that our proposed method outperforms the existing techniques on all the three datasets thanks to its comprehensive feature extraction and modeling through our new oscillometric data representation and the hybrid deep learning framework.

Unlike techniques that rely on the initial assessment of some cardiovascular parameters such as MAP and the estimation of SBP and DBP based on this initial assessment using empirical coefficients [27], our method is coefficient-free and does not rely on initial estimation of MAP. Through many theoretical and experimental studies, it has been shown that these methods are highly sensitive to changes in BP waveforms, pulse pressures, and arterial compliance and therefore they need to be tuned for different age groups, in different health conditions, and under different settings [28].

There have been several studies on the use of raw OMW envelope for the estimation of BP [1-3, 7, 11, 15]. These techniques mainly suffer from high estimation errors given that they solely are based on the analysis of the amplitude of the oscillometric pluses and ignore the wealth of information hidden in their morphology.

Many of the existing techniques have also been validated on small datasets which limits their generalizability to real-world data. DBN-DNN [3] was recently proposed as a technique based on generating artificial features using resampling algorithms like bootstraps to generate more data and enhance BP estimation. However, the level of improvement achieved by resampling was minimal since the number of original oscillometric measurements was relatively low. Additionally, the approach was based on selecting the best DBN-DNN estimator and discarding the rest which may lead to the loss of important information. A solution to this problem was provided by using an AdaBoost ensemble technique [7]. Another issue with such deep learning models [3, 7, 9] is their sensitivity to their parameter initialization stage that can cause uncertainty in BP estimates.

Unlike methods that use extra information such as age and gender for BP estimation [7, 11], our technique was solely based on the analysis of oscillometric measurement. Fusing such information to our network could further enhance the achieved results. There exist also methods that are based on the measurement of other physiological signals such ECG and BCG for improved BP estimation [13, 23, 29, 30]. However, the measurement of extra physiological signals is more complex and may not be feasible in an oscillometric BP monitor.

Unlike those methods that use data samples of the same individuals in training and test sets [12, 18], we ensured that our training-validation and test sets are consisted of data from different individuals and therefore our reported results better demonstrate the generalizability of our approach to unseen data. This is especially important when evaluating medical systems that work based on the analysis of physiological signals, as physiological characteristics may vary significantly between different individuals.

Unlike most of the existing techniques that are based on the analysis of expert-engineered features extracted from the oscillometric waveform [3, 7, 9, 11, 12, 18], our method provides an end-to-end deep learning framework that does not rely on expert-engineered features. Our proposed method automatically extracts the best representative features from the time pattern of the oscillometric waveforms and therefore, can capture complex input-output relationships that may not be visually observable.

*E. Limitations and Future Work*

While our developed method was validated on three different oscillometric datasets collected under different settings, it was not validated on patients with cardiovascular diseases. It has been shown that arterial stiffness, atrial fibrillation, and diabetes can impact the accuracy of oscillometric devices [31]. For example, individuals with varying levels of arterial stiffness may exhibit different hemodynamic relationships between BP and the measured oscillometric waveform and, as a result, our model may need to further be trained on data collected from such individuals.

Our datasets were also not uniformly distributed at different ranges of BP. Most of the recordings were around normal BP values with fewer samples at very low and high BPs. Having a uniformly distributed dataset can significantly improve the accuracy and reliability of our proposed method.

TABLE V
COMPARISON OF THE PROPOSED METHOD WITH CONVENTIONAL AND STATE-OF-THE-ART OSCILLOMETRIC BP MEASUREMENT METHODS. ERRORS ARE REPORTED IN MMHG.

| Methods | SBP | | DBP | |
|---|---|---|---|---|
| | SDE | ME | SDE | ME |
| Dataset1 | | | | |
| MAA [1] | 9.0 | 1.0 | 7.4 | 1.1 |
| FFNN [2] | 7.6 | -0.3 | 6.7 | -1.7 |
| SVR [3] | 7.1 | -0.0 | 6.0 | -0.2 |
| DBN-DNN [3] | 6.3 | 0.0 | 5.2 | 0.0 |
| DBN-DNN Ensemble [7] | 5.7 | -0.0 | 4.6 | -0.0 |
| DBR [9] | 5.9 | -0.5 | 4.8 | 0.2 |
| DBMDS [11] | 5.3 | 0.1 | 4.3 | -0.1 |
| **Proposed Method** | 2.4 | 0.0 | 2.2 | 0.0 |
| Dataset2 | | | | |
| MAA [12] | 13.3 | 0.1 | 12.0 | 1.2 |
| FFNN [12] | 11.8 | 0.2 | 12.0 | 2.0 |
| SVR [12] | 12.1 | 9.0 | 12.4 | 0.2 |
| LSTM-RNN [12] | 5.9 | -1.2 | 8.8 | 1.8 |
| DBN-DNN-BB [18] | 2.9 | 0.4 | 5.6 | -1.0 |
| **Proposed Method** | 3.6 | 0.1 | 3.4 | 0.1 |
| Dataset3 | | | | |
| PTT [13] | 5.8 | 3.0 | 4.5 | 2.3 |
| Ratio-Independent Modeling [16] | 5.8 | 0.0 | 5.9 | -1.7 |
| Bayesian Fusion [22] | 5.2 | -0.2 | 3.6 | -0.3 |
| **Proposed Method** | 3.7 | 0.2 | 2.2 | -0.0 |

To effectively capture the morphological and temporal relationships between the OMW and the BP, a hybrid CNN-RNN model was proposed. Techniques based on the attention mechanism and class activation maps can be adopted in a future work to further enhance the model interpretability and performance.

## IV. CONCLUSIONS

The relatively high BP estimation errors of the existing oscillometric algorithms prompt the inefficiency of the current oscillometric data representation techniques and suggests the need for further research in characterizing the oscillometric measurements. In addition, the current standards for BP estimation appears to be somewhat lax as, for example, the recommended ME and SDE by ANSI/AAMI are <5mmHg and <8mmHg, respectively. Here, to further improve the accuracy of oscillometric BP estimates, we proposed a new oscillometric data structure that can fully capture the OMW time and cuff pressure dependencies. A deep CNN-RNN model was developed to automatically extract the OMW informative features from the new data structure and estimate BP. The proposed method was validated on three different oscillometric datasets collected from upper arms and wrists of 245 healthy individuals under different settings. Our model resulted in a substantially lower ME and SDE than the current standard recommendations (ME<0.22 mmHg and SDE<3.72 mmHg) and outperformed state-of-the-art techniques. Future work will involve the validation of the proposed method on different patient populations including those with arterial stiffness, atrial fibrillation, and diabetes.